\documentclass[manuscript,screen,nonacm]{acmart}
\usepackage[utf8]{inputenc}
\AtBeginDocument{%
  }

\usepackage{pgfplots}
\pgfplotsset{compat=1.18}
\usepgfplotslibrary{groupplots}

\usepackage{listings}
\setcopyright{acmlicensed}
\copyrightyear{2018}
\acmYear{2026}
\acmDOI{XXXXXXX.XXXXXXX}
\acmConference[Conference acronym 'XX]{Make sure to enter the correct
  conference title from your rights confirmation email}{June 03--05,
  2018}{Woodstock, NY}
\acmISBN{978-1-4503-XXXX-X/2018/06}

\begin{document}

\title{ASTRA - Agentic System for Ticket Resolution and Analysis}

\author{Shashidhar Reddy Javaji}
\authornote{ Contributed equally to this research.}
\authornote{Work done while interning at Nokia Bell Labs.}
\email{sjavaji@stevens.edu}
\affiliation{%
  \institution{Stevens Institute of Technology}
  \city{Hoboken}
  \state{New Jersey}
  \country{USA}
}

\author{Mohamed Trabelsi}
\authornotemark[1]
\email{mohamed.trabelsi@nokia-bell-labs.com}
\affiliation{%
  \institution{Nokia Bell Labs}
  \city{Murray Hill}
  \state{New Jersey}
  \country{USA}
}

\author{Jin Cao}
\affiliation{%
  \institution{Nokia Bell Labs}
  \city{Murray Hill}
  \state{New Jersey}
  \country{USA}
}

\author{Huseyin Uzunalioglu}
\affiliation{%
  \institution{Nokia Bell Labs}
  \city{Murray Hill}
  \state{New Jersey}
  \country{USA}
}









\renewcommand{\shortauthors}{Javaji et al.}

\begin{abstract}
Technical operations teams resolve large volumes of incidents and support tickets by synthesizing fragmented evidence from ticket text, historical cases, system logs, and technical documentation. This process is slow and error-prone, and existing automation often relies on monolithic, single-shot generation that lacks explicit evidence modeling and provenance, making outputs difficult to verify when critical signals are sparse and scattered across sources. We propose ASTRA, an agentic system for ticket resolution and analysis in which a central orchestrator coordinates three specialist information-gathering agents and drives a judge--orchestrator refinement loop to produce evidence-backed troubleshooting reports. A \emph{TicketSimilarityAgent} retrieves relevant historical precedents through a two-phase dense-retrieval and LLM-reranking pipeline; a \emph{LogAgent} distills hundreds of thousands of log lines into structured, quote-grounded findings via deterministic filtering followed by a constrained LLM analysis; and a \emph{DomainKnowledgeAgent} retrieves relevant documentation and technical knowledge via the Model Context Protocol (MCP). The three agents' outputs are then transformed into an explicit claim--evidence intermediate representation that links every claim to a verbatim source passage, annotates it with a support level, and prevents cross-attribution between agents. A \emph{JudgeAgent} scores the resulting report on five rubric criteria, and an \emph{OrchestratorAgent} translates low-scoring metrics into a small set of targeted follow-up queries; the loop runs for a bounded number of rounds. Evaluated on 987 real-world telecom fault tickets across seven product lines, ASTRA achieves a mean quality score of 4.13/5.0 across five evaluation dimensions, with 59.9\% of reports correctly identifying the fault area at the component-family level or better. Presentation quality is consistently high (Relevance~4.88, Clarity~4.94), and the claim--evidence architecture limits fabricated technical details to under 3\% of error cases. Stratification by fault type reveals that hardware faults remain substantially harder than software or configuration faults (Cohen's $d$=0.80), pointing to a fundamental limitation of text-based evidence channels for hardware fault diagnosis.
\end{abstract}





\maketitle

\section{Introduction}
Technical operations teams resolve large volumes of incident and support tickets under tight time constraints. A single ticket often requires synthesizing \emph{heterogeneous, partially redundant, and frequently noisy} evidence from (i) the ticket narrative and metadata, (ii) historical tickets and prior resolutions, (iii) system logs / telemetry, and (iv) documentation and troubleshooting guides. In practice, engineers iteratively form hypotheses, gather targeted evidence, validate (or refute) competing explanations, and finally produce a resolution report that must be both \emph{actionable} and \emph{auditable}-especially in availability-critical domains such as telecommunications \cite{trabelsi_teledoctr:_2026}.

A substantial body of research has targeted \emph{individual components} of the ticket lifecycle. On the ``front end'', incident and ticket \emph{triage} systems aim to route or categorize tickets efficiently and consistently; large language models have been shown to provide accurate and interpretable incident triage in production-oriented settings \cite{wang_large_2024}, and multi-agent formulations can emulate multi-role workflows during assignment \cite{yu_triangle:_2025}. In parallel, work on \emph{ticket escalation} addresses operational realities where decisions must be topic-aware and dynamic as tickets evolve \cite{liu_tickit:_2025}. On the ``back end'', LLM-augmented methods for \emph{ticket aggregation} reduce resolution effort by consolidating related reports and supporting low-cost defect handling \cite{sun_llm-augmented_2025}. While valuable, these lines of work typically optimize isolated stages rather than supporting evidence-driven, end-to-end troubleshooting.

Recent telecom-focused systems have taken an important step toward integration. TeleDoCTR \cite{trabelsi_teledoctr:_2026} combines domain-specific ranking with generation to automate routing, similar-ticket retrieval, and fault analysis report drafting outlining the issue, root cause, and potential solutions for telecom troubleshooting. However, end-to-end ticket resolution remains challenging when evidence is distributed across sources with differing reliability and when the system must decide \emph{what to investigate next} to close gaps. More broadly, realistic agent benchmarks suggest that robust IT automation is still far from solved where ITBench reports low end-to-end success rates for state-of-the-art agents on  Site Reliability Engineering (SRE)-style scenarios \cite{jha_itbench:_2025}, and similarly low end-to-end success persists in realistic web and tool-use environments and software engineering workflows \cite{zhou_webarena:_2023,mialon_gaia:_2024,jimenez_swe-bench:_2024,yang_swe-agent:_2024}. These observations point to a core challenge beyond fluency-\emph{controllability, verification, and evidence-grounded iteration} in complex, tool-heavy workflows.

In response, the community has increasingly explored \emph{agentic} and \emph{tool-augmented} paradigms for diagnosis and remediation \cite{parisi_talm:_2022,wu_autogen:_2024,li_camel:_2023,hong_metagpt:_2024}. LLM-based root cause analysis (RCA) systems can automate incident analysis by collecting diagnostic signals and generating explanatory narratives \cite{chen_automatic_2024}, and LLMs can also improve investigation efficiency by recommending domain-specific diagnostic queries grounded in historical incidents \cite{jiang_xpert:_2024}. LLexus \cite{las-casas_llexus:_2024} further frames incident handling as the execution of troubleshooting guides by an agent that converts procedural knowledge into actionable plans. At a systems level, STRATUS \cite{chen_stratus:_2025} demonstrates how specialized agents can be organized under explicit control logic and safety constraints for iterative mitigation.

This paper builds on that trajectory and argues that ticket troubleshooting is best approached as orchestrated multi-agent evidence synthesis. We propose an agentic telecom-related ticket troubleshooting framework in which a single orchestrator coordinates specialized agents aligned with key evidence channels: a \emph{TicketSimilarityAgent} retrieves high-utility historical precedents; a \emph{LogAgent} performs structured extraction over large log streams to surface diagnostically ``golden'' signals; and a \emph{DomainKnowledgeAgent} grounds reasoning in canonical knowledge by querying domain-specific documentation via the Model Context Protocol (MCP). Each agent produces bounded artifacts intended for downstream verification rather than unconstrained narrative text. Because troubleshooting often spans long horizons and repeated interactions, we explicitly account for long-term memory and state management mechanisms inspired by recent agent memory systems \cite{packer_memgpt:_2024,xu_-mem:_2025,park_generative_2023}. The orchestrator then constructs an explicit \emph{claim--evidence intermediate representation}: candidate claims paired with provenance-linked evidence snippets from one or more sources, annotated with support levels and cross-source consistency signals. This intermediate layer enables controllable report generation, targeted follow-up queries, and auditable traceability.

A key design goal is to avoid the brittleness of monolithic end-to-end generation by using iterative refinement with evidence attribution and explicit planning. RARR \cite{gao_rarr:_2023} shows how generation can be post-edited to add attribution and remove unsupported content, and Self-RAG \cite{asai_self-rag:_2024} motivates retrieval-and-critique behaviors \cite{wang_plan-and-solve_2023} that determine when retrieval is needed and how evidence should shape generation. We operationalize these ideas via a \emph{judge-orchestrator loop}: a judge agent evaluates draft reports against task-specific criteria (e.g., root-cause plausibility and identification quality), flags weak or under-supported sections, and triggers targeted follow-ups to the appropriate specialist agent(s). To reduce unsupported conclusions, we incorporate explicit verification behaviors shown to mitigate hallucinations \cite{dhuliawala_chain--verification_2024,manakul_selfcheckgpt:_2023}. Because LLM-based evaluation can be inconsistent, we follow reliability-oriented guidance for ``LLM-as-a-Judge'' systems \cite{gu_survey_2026,zheng_judging_2023}. For retrieval-augmented components, we incorporate reference-free diagnostic signals inspired by RAG evaluation frameworks such as RAGAs \cite{es_ragas:_2024}.

In summary, we make the following contributions:
\begin{itemize}
  \item \textbf{An end-to-end multi-agent troubleshooting architecture} that coordinates specialized agents over historical tickets, logs/telemetry, and documentation to produce evidence-backed fault analysis (FA) reports outlining the issue, root cause, and potential solutions.
  \item \textbf{A claim--evidence intermediate representation} with provenance links and cross-source consistency checks to improve traceability and support controlled refinement.
  \item \textbf{A judge--orchestrator refinement policy} with targeted follow-ups and bounded rounds, designed to reduce unsupported conclusions while preserving operational usefulness.
  \item \textbf{An evaluation protocol} that measures report quality alongside evidence grounding and retrieval faithfulness, informed by realistic agent benchmarking\cite{jha_itbench:_2025,es_ragas:_2024}.
\end{itemize}

\section{Related Work}
\subsection{Ticket Troubleshooting}
Operational troubleshooting has recently been reframed as a sequence of ticket-centric tasks---routing/triage, retrieval of similar historical cases, and generating resolution artifacts. Early LLM-based systems and studies show that language models can support interpretable incident triage decisions \cite{wang_large_2024}, while other lines of work improve operator productivity by recommending better diagnostic queries \cite{jiang_xpert:_2024,jiang_xpert:_2024-1}. In parallel, ticket workflow automation has been decomposed into specialized sub-problems such as ticket summarization \cite{absformer}, escalation \cite{liu_tickit:_2025} and aggregation/deduplication for faster defect resolution \cite{sun_llm-augmented_2025,sun_llm-augmented_2025-1}. Complementary efforts profile and structure ticket content itself (e.g., hierarchical fault profiling) to enable downstream analytics and decision-making \cite{huang_faultprofit:_2024}.

Beyond isolated subtasks, recent systems move toward more end-to-end troubleshooting pipelines. TeleDoCTR \cite{trabelsi_teledoctr:_2026} explicitly unifies classification, retrieval, and report generation for telecom ticket resolution. In cloud incident settings, LLMs have been used to generate root-cause analyses and mitigation steps \cite{chen_automatic_2024,chen_automatic_2024-1,ahmed_recommending_2023}, and agentic approaches such as RCAgent \cite{wang_rcagent:_2024} emphasize tool-augmented interaction for collecting evidence during RCA. Industrial on-call automation has also begun exploring explicit multi-agent collaboration as a design pattern \cite{DBLP:conf/kbse/FuZCWZRZWSLLZ25}. However, these systems still often rely on brittle orchestration (fixed stages or Standard Operating Procedure (SOP) templates) and do not consistently expose controllable intermediate representations (e.g., explicit hypotheses, evidence sets, and verification outcomes) that operators can audit.

A growing body of work argues that multi-agent decomposition can reduce cognitive load and improve robustness in high-stakes operations. Triangle \cite{yu_triangle:_2025} introduces multi-agent incident triage to route incidents more effectively, and STRATUS \cite{chen_stratus:_2025} explores multi-agent autonomy for reliability engineering in modern clouds. More explicitly RCA-focused multi-agent methods incorporate structured workflows and coordination mechanisms, e.g., SOP-enhanced multi-agent RCA \cite{pei_flow--action:_2025}, knowledge-base driven troubleshooting workflow coevolution \cite{shi_flowxpert:_2025}, and blockchain-inspired voting/termination controls for microservices RCA \cite{zhang_mabc:_2024}. Even when these approaches improve automation accuracy, they still highlight persistent gaps: (i) \emph{context fragility} when evidence is distributed across tickets, logs, and documentations, (ii) \emph{error propagation} across long multi-step reasoning chains, and (iii) limited \emph{controllability} over what the system asserts vs.\ what it can justify. Recent multi-agent-RCA work makes these failure modes explicit and motivates specialized agents (retrieval/validation) to counter hallucinations and context-switching errors \cite{fu_leveraging_2025}. These trends collectively motivate a multi-agent fault analysis framework that is not just task-compositional, but also \emph{evidence-grounded and verifiable} by construction.

\subsection{Agentic Systems}
The broader agentic systems literature provides the scaffolding needed to operationalize such decomposition. Tool augmentation is a foundational capability for agents acting over external state, with TALM \cite{parisi_talm:_2022} demonstrating how non-differentiable tools can be integrated via demonstration and iterative improvement. Planning-centric prompting methods (e.g., Plan-and-Solve) further show that explicit decomposition can reduce missing-step and reasoning errors in multi-step tasks \cite{wang_plan-and-solve_2023}. On the coordination side, multi-agent frameworks range from communication-driven role play \cite{li_camel:_2023} and agent conversation abstractions \cite{wu_autogen:_2024} to more structured meta-programming approaches for collaborative software production \cite{hong_metagpt:_2024} and reflective collaboration mechanisms that emphasize iterative critique and revision \cite{bo_reflective_2024}.

A recurring challenge in real deployments is that long-horizon tasks stress both memory and environment interaction. Systems such as MemGPT \cite{packer_memgpt:_2024} treat memory as a managed hierarchy to extend effective context beyond window limits, while A-Mem \cite{xu_-mem:_2025} proposes agentic memory mechanisms tailored to LLM agents. Meanwhile, realistic evaluation environments and benchmarks expose how difficult it remains to build robust autonomous agents: WebArena \cite{zhou_webarena:_2023} provides a grounded web environment for agent interaction, GAIA \cite{mialon_gaia:_2024} targets general AI assistants, and software-engineering benchmarks highlight brittleness at the repository level \cite{jimenez_swe-bench:_2024} even when agent-computer interfaces are introduced \cite{yang_swe-agent:_2024}. Multi-agent issue-resolution systems such as MAGIS \cite{tao_magis:_2024} attempt to address these gaps via specialized roles (manager/developer/QA) but still achieve modest solve rates, underscoring the difficulty of reliable orchestration. Similarly, ITBench \cite{jha_itbench:_2025} shows that state-of-the-art agents resolve only a minority of real-world IT automation scenarios, indicating substantial headroom for reliability-centric agent design.

Taken together, these results suggest that merely adding agents is insufficient to address operational fault analysis demands: (i) stable shared state (memory), (ii) controllable decomposition, and (iii) grounded interaction with tools and evidence. This directly motivates frameworks that explicitly separate \emph{analysis claims} from \emph{supporting evidence} and provide systematic verification hooks, rather than relying on implicit end-to-end generation.

\subsection{Claim Evidence}
A key missing ingredient in many operational RCA pipelines is a principled claim--evidence interface that supports attribution, verification, and post-hoc auditing \cite{javaji-etal-2025-ai}. In text generation settings, RARR \cite{gao_rarr:_2023} demonstrates how to retrofit attribution by explicitly researching and revising model outputs. Retrieval-augmented generation methods increasingly incorporate self-critique and reflection (e.g., Self-RAG \cite{asai_self-rag:_2024}) to decide when to retrieve, how to critique, and how to revise generations. At the system level, evaluation frameworks such as RAGAs provide reference-free metrics that separately assess retrieval quality and faithfulness of the generated answer \cite{es_ragas:_2024}. For hallucination reduction, Chain-of-Verification (CoVe) \cite{dhuliawala_chain--verification_2024} formalizes a staged procedure where models draft, generate verification questions, answer them independently, and then produce a revised response; complementary black-box approaches such as SelfCheckGPT \cite{manakul_selfcheckgpt:_2023} detect inconsistencies by sampling multiple generations.

Reliability also depends on \emph{how} we evaluate complex reasoning outputs. Recent work on LLM-as-a-Judge analyzes judge design, bias, and reliability issues, and proposes methodological standardization for using LLMs as evaluators \cite{gu_survey_2026}. Benchmarks like MT-Bench/Chatbot Arena have been used to study judge behavior and its failure modes \cite{zheng_judging_2023}. In claim verification specifically, surveys document the shift toward LLM- and RAG-based verification pipelines, including common retrieval, prompting, and fine-tuning patterns \cite{dmonte_claim_2025}. Fine-grained, explainable verification systems such as ClaimVer \cite{dammu_claimver:_2024} emphasize claim-level verification with evidence attribution and human-centric explanations. Newer datasets push beyond single-hop text-only verification: MMCV \cite{wang_piecing_2025} targets multi-hop \emph{multimodal} claim verification across text, images, and tables, while CliniFact \cite{zhang_dataset_2025} provides a clinical research claim benchmark that exposes the gap between discriminative and generative verification performance. Finally, system demonstrations like CEDAR \cite{jayasekara_demonstrating_2025} highlight cost-efficient, data-driven verification pipelines that are closer to operational constraints.

Overall, the evidence-grounded generation and verification literature indicates that robust systems require explicit intermediate artifacts (claims, evidence, attribution, verification results) plus reliable evaluation. Yet, many ticket/incident systems still produce narrative fault analysis reports without explicit, checkable claim--evidence structure. This motivates a multi-agent fault analysis framework that (i) decomposes troubleshooting into controllable sub-steps, (ii) maintains and curates context over long horizons, and (iii) generates reports as verifiable sets of claims with attached evidence and validation outcomes, aligning operational RCA with best practices in grounded generation and claim verification.

\section{Problem}

We study the problem of \emph{automated fault analysis report generation} for technical support tickets in telecom infrastructure. Formally, given a target ticket $T$ consisting of structured metadata (product, feature, software build) and unstructured text (title and problem description), along with references to associated log files, the goal is to produce a fault analysis report $R = (\text{identification}, \text{resolution})$ where the \emph{identification} section describes how the fault was detected and what caused it, and the \emph{resolution} section describes the fix that should be applied.

Generating this report requires synthesizing evidence from three heterogeneous external channels alongside the ticket itself: (i) a corpus of historical tickets and their resolved fault analyses, from which relevant precedents can be retrieved; (ii) raw system logs collected from the affected network node, typically comprising hundreds of thousands of log lines drawn from separate system and runtime log streams; and (iii) a body of technical documentation including alarm definitions, configuration guides, and known-issue references. Each channel contributes a different kind of signal---pattern-based precedent, concrete runtime evidence, and normative specification, respectively---and a complete analysis ideally draws on all three.

Several factors make this a non-trivial problem:

\textbf{Evidence quality and coverage are uneven.} Similar historical tickets may describe related but not identical fault scenarios. Log files are large and noisy: critical error indicators are intermixed with routine system events, and the relevant lines may be scattered rather than contiguous. Documentation may cover the component in question but not the specific configuration or version deployed.

\textbf{Sources must be cross-validated, not just aggregated.} A root cause hypothesis that emerges from a historically similar ticket may be inconsistent with what the logs for the current ticket actually show. Conversely, a log-derived signal may be ambiguous without documentation context. Simply concatenating source outputs and feeding them to a generation model risks producing a report that blends conflicting signals without acknowledging the conflict.

\textbf{Reports must be auditable, not just fluent.} In availability-critical environments, a report that \emph{sounds} confident but cannot be verified against the evidence it was derived from is often worse than no report at all. Operators need to know which parts of an analysis are strongly grounded and which rely on weaker or indirect signals, so that they can triage the report's conclusions appropriately.

\textbf{The long-context challenge.} Log files attached to tickets can span tens of thousands of lines. Passing the full log stream as LLM context is expensive and tends to dilute the model's attention away from the relatively small number of lines that are diagnostically informative. A principled selection strategy is needed that is both accurate (does not discard critical lines) and conservative (does not hallucinate evidence that is not present).

We further restrict the generation target to a \emph{structured, evidence-backed} report in which every substantive claim is linked to the source that supports it, and claims are explicitly annotated with a support level so that downstream evaluation can distinguish well-grounded findings from weaker inferences. This framing enables our system to generate outputs that are both practically useful for engineers and amenable to automated evaluation.

\section{ASTRA}

\begin{figure}[h]
    \centering
    \includegraphics[width=0.79\linewidth]{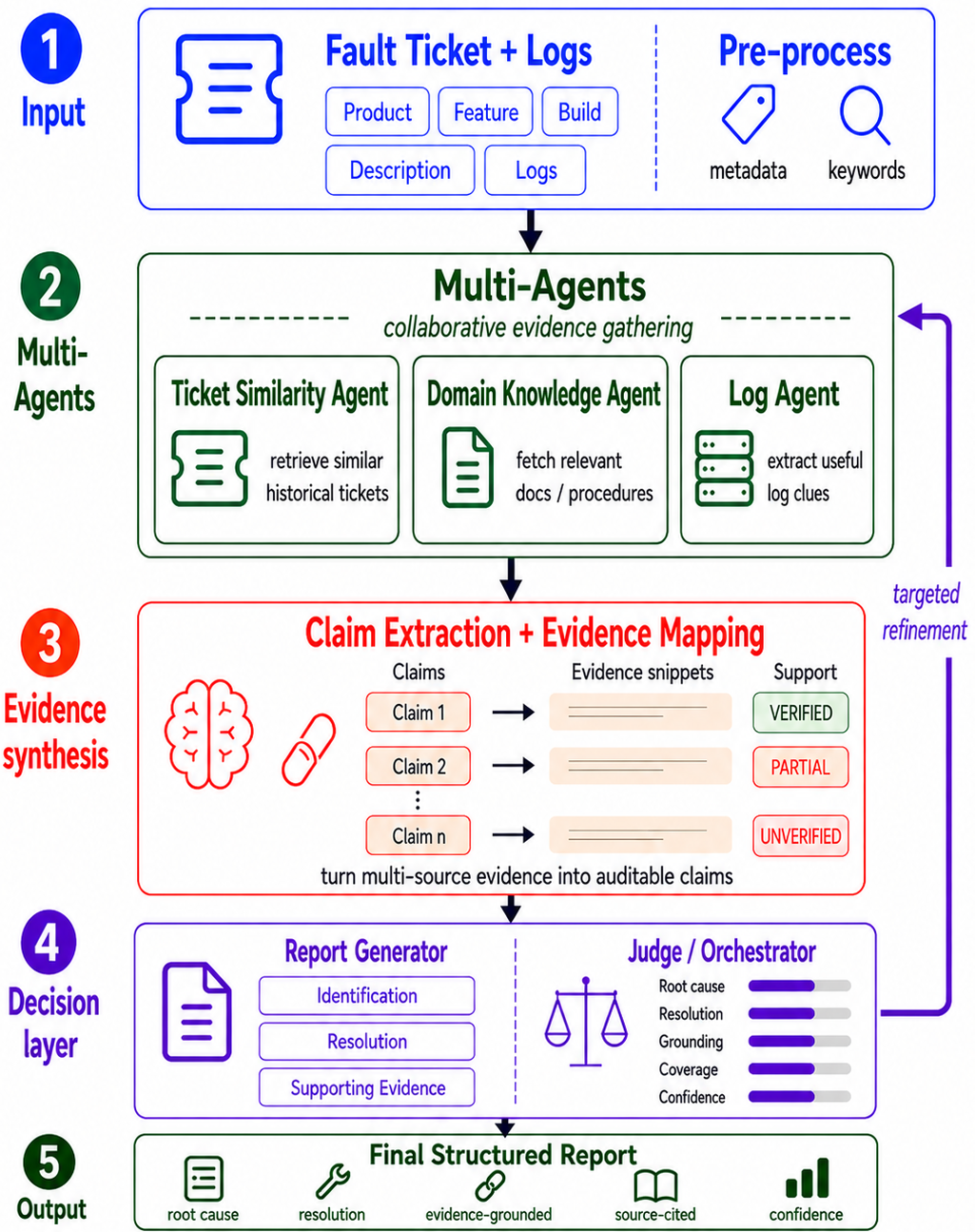}
   \caption{ASTRA overview. Starting from a fault ticket and preserved logs, ASTRA invokes three specialized agents (TicketSimilarityAgent, DomainKnowledgeAgent, and LogAgent) to gather complementary evidence. Their outputs are converted into a claim--evidence intermediate representation with support labels, which is then used to generate a structured report. A Judge/Orchestrator scores each draft against rubric criteria and issues targeted follow-up queries to the most relevant agents for a bounded number of refinement rounds before producing the final report.}
    \label{fig:astra_architecture}
\end{figure}

We describe ASTRA (\textbf{A}gentic \textbf{S}ystem for \textbf{T}icket \textbf{R}esolution and \textbf{A}nalysis), a multi-agent framework for automated fault analysis report generation. ASTRA coordinates three specialist information-gathering agents, a structured claim--evidence construction pipeline, and a judge--orchestrator refinement loop (Figure~\ref{fig:astra_architecture}).

The central design decision in ASTRA is to separate \emph{information gathering} from \emph{report generation}, with an explicit intermediate representation sitting between them. Rather than directly prompting a model to produce a fault analysis from raw ticket context, ASTRA first collects and structures evidence through specialized agents, constructs a claim--evidence mapping that annotates each finding with its source and confidence level, and then generates the report from this structured representation. This decoupling makes intermediate outputs inspectable, reduces the surface area for hallucination, and allows the refinement loop to identify and address specific gaps rather than revising the report holistically.

\subsection{Agent Architecture}

ASTRA operates three specialist information-gathering agents, each aligned with one of the external evidence channels described in Section~3. The agents are implemented on top of AutoGen's agent communication framework \cite{wu_autogen:_2024}, and each exposes a uniform \texttt{run(task)} interface so the orchestrator can query them identically during initial evidence collection and in follow-up rounds.

\textbf{TicketSimilarityAgent.} This agent retrieves historical tickets that are semantically similar to the target and returns formatted summaries of their fault analyses and resolutions as its output artifact. Internally it uses a two-phase retrieval strategy (full reranker prompts in Appendix~\ref{prompt:ticketsimilarity}). In the first phase, dense vector search over a database of approximately 130,000 enhanced historical tickets fetches a broad candidate set of up to 80 tickets. Our dense retriever is an MPNet-initialized domain-specific ranker trained with large-batch contrastive learning (in-batch negatives; GradCache \cite{GaoZHC21} for memory efficiency), which yields substantially stronger telecom retrieval than vanilla MPNet. Ticket representations in this database have been enriched with LLM-generated expanded descriptions, acronym expansions, and component context during an offline preprocessing step, which improves query coverage for abbreviated or domain-specific fault descriptions. In the second phase, an LLM-based reranker proceeds in two passes. The first pass rescores the 80 candidates in four batches of 20 and selects the top-5 from each batch (20 intermediate results); the second pass rescores those 20 and selects the final top-$k$ (we use $k=5$), discarding any candidate whose reranker-adjusted similarity score falls below 0.65. For the top two ranked results, the full investigation trail recorded by the technical analysis (TA) team (where available) is appended to the ticket representation, giving downstream stages richer context about how engineers previously diagnosed similar faults. As described in the execution schedule below, this two-phase pipeline is run once per ticket as a shared pre-fetch whose output is consumed both by this agent and by the LogAgent.

\textbf{LogAgent.} Software logs play a central role in diagnosis and automation tasks, as they often constitute the primary source of information capturing software runtime behavior. These collected logs are leveraged in a wide range of log mining applications, including anomaly detection \cite{loggpt,trabelsi2025anomaly}, failure prediction \cite{ChenYLZGXDZDXLK19}, and failure diagnosis \cite{JiaCYLMX17}. LogAgent analyzes the raw system and runtime log files attached to a ticket and returns a structured analysis containing diagnostic findings, a causal chain, and a root cause category. ASTRA processes log files in full without truncation, with individual log streams that can contain hundreds of thousands of lines. The pipeline proceeds in four stages:

\emph{Loading}: both syslog and runtime log parquet files are read into memory in full, with severity labels derived from the feature name column (\texttt{ERR}, \texttt{WRN}, \texttt{CRIT}, etc.).

\emph{Deterministic filtering}: a five-level filter identifies diagnostically relevant lines without any LLM involvement. Level~1 selects lines whose severity label indicates an error or warning. Level~2 matches a curated set of error and fault keywords in the message text (alarm, fault, fail, exception, crash, timeout, etc.). Level~3 captures state-transition events that are diagnostically meaningful even at informational severity (restart, deregister, handover failure, sync loss, etc.). Level~4 targets component-level anomaly indicators (core dump, stack trace, watchdog, queue overflow, etc.). Level~5 extracts keywords from the ticket's own title and description \emph{and} from the titles and descriptions of the pre-fetched similar tickets, using them to catch relevant lines specific to the reported component or feature that generic patterns would miss. Results from all five levels are unioned and deduplicated by line index. A noise suppressor then identifies high-volume background components that account for more than 25\% of diagnostic lines but contain no fault-severity keywords (e.g., transport-layer protocol chatter) and replaces their individual lines with a single metadata annotation so they do not anchor the LLM's attention.

\emph{Context expansion}: each retained diagnostic line is expanded to include $\pm5$ surrounding lines from the original log stream, providing the LLM the narrative flow around each event. If the combined set exceeds an upper limit, DRAIN3 \cite{he2017drain} template clustering is applied to deduplicate structurally similar lower-severity lines while preserving all distinct error and warning messages. The post-filtering excerpt is capped at 2,000 lines before being passed to the LLM; 84\% of tickets in our evaluation saturate this limit.

\emph{Constrained LLM analysis}: the resulting log excerpt is passed to the LLM together with the ticket's enhanced description, acronym expansions, and component context, and---crucially---the shared pre-fetch data drives two distinct behaviours here. First, as noted above, keywords from similar ticket titles and descriptions augment Level~5 filtering to surface log lines specific to the fault's component. Second, the resolution summaries and root cause labels from the pre-fetched tickets are formatted with rank labels (HIGH MATCH for $\geq$1.0 reranker score, GOOD MATCH for $\geq$0.9, and POSSIBLE MATCH for scores below 0.9) and appended to the LLM prompt as a labelled reference block (``SIMILAR RESOLVED HISTORICAL TICKETS''), with TA investigation trails included for the top two ranked matches only---TA trails can extend to several hundred lines of debugging notes, so including them for all five candidates would inflate the prompt without proportional diagnostic benefit, given the lower transferability of lower-ranked resolutions; the LLM is instructed to use these only to cross-validate its log findings---not to copy them verbatim (Appendix~\ref{prompt:logdigest}). The output is a structured JSON object with a causal narrative, the supporting log evidence, and a root cause category selected from a fixed taxonomy of 53 categories derived from the historical ticket database.

\textbf{DomainKnowledgeAgent.} This agent retrieves content from technical documentation and returns findings relevant to fault identification and resolution (full prompt in Appendix~\ref{prompt:domainknowledge}). It queries a domain knowledge base via the
MCP, which provides access to alarm specifications, configuration guides, and known-issue references.

The query issued to the MCP backend is constructed to be as targeted as possible. It incorporates the ticket title, any fault identifier codes (FIDs) extracted from the title via regex, and---crucially---up to 1,500 characters of TicketSimilarityAgent output and 2,000 characters of LogAgent output. This means the DomainKnowledgeAgent can issue queries like ``what configuration parameter causes this alarm given that the log shows X'' rather than a generic alarm lookup. The agent's system message instructs it to distinguish identification-relevant findings (symptom meanings, diagnostic procedures, alarm definitions) from resolution-relevant findings (configuration changes, workarounds, known-issue fixes) in its output, which aligns with the two-section structure of the fault analysis report.

\textbf{Execution schedule.} Execution proceeds in three steps.
First, a single synchronous pre-fetch runs the two-phase retrieval pipeline
and stores the results as raw structured dictionaries---one per retrieved
ticket---carrying root cause labels, resolution text, reranker scores, and TA
investigation trails for the top two results. These dictionaries are shared
between both Phase~2a agents: the TicketSimilarityAgent formats them into
the human-readable summaries that constitute its output artifact, while the
LogAgent receives them at construction time to drive Level~5 keyword
filtering and the reference block injected into its LLM prompt. Because both
agents consume the same pre-fetched data, they are mutually input-independent
at the start of Phase~2a and execute concurrently via \texttt{asyncio.gather}.
Second, once Phase~2a completes, the DomainKnowledgeAgent runs in Phase~2b; the
orchestrator embeds up to 1,500 characters of TicketSimilarityAgent output
and 2,000 characters of LogAgent output directly into its task string, allowing
it to issue documentation queries conditioned on both agents' findings. Third,
the three agent outputs are passed to the claim--evidence construction pipeline
before final report generation and refinement.

\subsection{Claim--Evidence Intermediate Representation}

Once the three agents have produced their outputs, ASTRA constructs a structured intermediate representation before generating the fault analysis report. This proceeds in two steps.

\textbf{Claim extraction.} Each agent's output is processed independently to extract a set of detailed, self-contained diagnostic claims; the extraction prompt is shown below. The extraction model is instructed to produce 5--10 claims per agent, annotating each with a \texttt{[CAUSAL]} prefix where the claim describes a root cause or direct trigger, and a \texttt{[SYMPTOMATIC]} prefix where it describes an observed downstream effect. When the distinction is unclear, no label is applied as we found that forcing uncertain labeling introduced more noise than leaving claims unlabeled.

A key constraint is that claims must be grounded strictly in the agent's own output. Agent-specific system messages explicitly prohibit cross-attribution: TicketSimilarityAgent claims must describe what retrieved historical tickets show (citing their problem report identifiers and similarity scores), LogAgent claims must reference only entries visible in the log analysis output, and DomainKnowledgeAgent claims must cite only content appearing in the retrieved documents. This discipline is important in practice because the same log line or historical resolution could appear plausibly relevant to all three agents, and without explicit attribution boundaries the model would tend to repeat the same finding in all three claim sets with slightly different framing, producing a misleadingly inflated pool of evidence.

\begin{lstlisting}[basicstyle=\footnotesize\ttfamily,breaklines=true,keepspaces=true]
Claim Extraction Prompt
-----------------------------------------------------------------------
Extract detailed, information-rich claims for a telecom fault analysis report.
Focus on two sections:
  1. IDENTIFICATION -- How was the problem detected and diagnosed?
Include: alarm IDs, log messages, error patterns, component names, root cause.
  2. RESOLUTION -- How was the problem resolved?
Include: exact fixes, configuration parameters (before/after), commands, workarounds, and verification steps.

CRITICAL GROUNDING RULE:
- Extract ONLY information explicitly stated in the AGENT OUTPUT provided.
- The target ticket context is background only -- do NOT use it to invent claims.
- Do NOT cross-attribute between agents.
- When in doubt, omit.
CAUSAL vs SYMPTOMATIC LABELLING (use where unambiguous):
  [CAUSAL]     -- direct trigger or root cause
  [SYMPTOMATIC] -- downstream effect or observed symptom
  (omit label if unsure)
CLAIM QUALITY:
- Each claim must be self-contained with specific technical details verbatim from output.
- Broader observational findings are valuable even without a specific ID.
- Negative findings (e.g., "no further alarms after fix") should be included.

Output: one claim per line. Extract 5-10 claims covering identification and resolution.
\end{lstlisting}

\textbf{Evidence mapping.} For each extracted claim, ASTRA uses the LLM to locate the specific passage in the originating agent output that most directly supports the claim (shown below). The extraction prompt instructs the model to quote the text verbatim, prefer passages containing exact identifiers (error codes, module names, parameter values) over paraphrases, and include one or two surrounding sentences for context where they add diagnostic value. The extracted evidence is validated by checking that at least 60\% of its words appear in the original agent output; this is a sourcing fidelity check---it detects fabricated technical identifiers (error codes, parameter names, module paths) that would score highly on semantic similarity but do not actually appear in the source---rather than a semantic retrieval step. Passages falling below this threshold are flagged as potentially hallucinated and replaced by a keyword-based fallback that scores sentence windows by keyword overlap. Each evidence item is annotated with a confidence level (high, medium, or low) and a status (supported or weak), and items meeting the confidence threshold are tagged as \texttt{[VERIFIED]} for the downstream generation step. Items that do not pass are retained but marked \texttt{[UNVERIFIED]} so the report generator can handle them conservatively.

\begin{lstlisting}[basicstyle=\footnotesize\ttfamily,breaklines=true,keepspaces=true]
Evidence Mapping Prompt
-----------------------------------------------------------------------
You are a precise evidence extraction assistant for telecom fault reports.
Find the EXACT text from the agent output that supports each claim.

CRITICAL RULES:
1. QUOTE DIRECTLY -- copy exact text, do NOT paraphrase or summarize.
2. MOST DIAGNOSTIC LINE FIRST -- prefer passages with error codes, fault IDs, or parameter values; do not skip observational evidence (e.g., load patterns, timing, state sequences) just because they lack a specific ID.
3. SURROUNDING CONTEXT -- include 1-2 sentences before/after if they add value.
4. NO INVENTION -- if no supporting text exists, return "No direct evidence found" and mark status as weak.

Return JSON:
{
  "claim":      "<original claim>",
  "evidence":   "<exact quoted text from agent output>",
  "source":     "<agent name>",
  "status":     "supported" | "weak",
  "missing":    "<if weak, what specific detail is absent>",
  "confidence": "high" | "medium" | "low"
}

Confidence:
  high   -- evidence directly and specifically supports the claim
  medium -- evidence is in the right area but lacks full specificity
  low    -- connection requires significant inference
\end{lstlisting}

\subsection{Report Generation and Refinement}

\textbf{Initial report generation.} The fault analysis report is generated from the claim--evidence mapping using a structured prompt that enforces a causal narrative format (see Appendix~\ref{prompt:report}). The prompt specifies that the \emph{identification} section must trace an event chain from the initial symptom through the diagnostic evidence to a confirmed or likely root cause, and the \emph{resolution} section must name the specific fix applied along with verification that the issue was resolved. Claims labeled \texttt{[VERIFIED]} are treated as factual; \texttt{[UNVERIFIED]} claims may be included but must be flagged with hedged language. If the TicketSimilarityAgent signals a potential root cause category from historical matches, this is injected as a low-weight background hint with explicit instructions not to use it to override log evidence.

\textbf{Judge evaluation.} After the initial report is generated, a JudgeAgent evaluates it against five criteria (see Appendix~\ref{prompt:judge}). The judge receives the ticket context, all three agent outputs, the claim--evidence mapping, and the draft report; it has no access to the ground-truth FA or TA investigation chain, scoring purely on internal consistency and evidence grounding. The five criteria are: \emph{Root Cause Identification} (does it name a specific mechanism and component?), \emph{Resolution Specificity} (are the fix steps concrete and grounded?), \emph{Evidence Grounding} (what fraction of specific technical identifiers in the report can be traced back to agent outputs?), \emph{Coverage and Completeness} (are all four expected dimensions---symptom, root cause, fix, and verification---present?), and \emph{Focus and Relevance} (does the report stay on topic or build its narrative around off-topic agent findings?). Each criterion is scored 1--5, and a score at or below 3 on any criterion triggers a revision. The threshold is intentionally set at 3 rather than the usual midpoint of 2 because based on our experiments with multiple tickets, a score of 3 in our rubric corresponds to a report that addresses the right topics but lacks grounding specificity---a condition that additional targeted queries can often remedy.

\textbf{Orchestrator planning and refinement loop.} When revision is needed, an OrchestratorAgent receives the judge's evaluation and plans a small set of targeted follow-up queries (typically two to four) directed at the agent or agents most likely to fill the identified gaps. For example, if the root cause identification scored 3 because the log analysis did not pinpoint the triggering component, the orchestrator issues a focused LogAgent query asking specifically about that component. Query deduplication is maintained across rounds so that the same query is never re-submitted in a subsequent iteration. The follow-up results are appended to the original evidence, and the report is revised against both the original claim--evidence mapping and the new findings. This loop runs for up to five rounds; per-round metric changes and convergence behavior are reported in Section~\ref{sec:refinement}.

\begin{lstlisting}[basicstyle=\footnotesize\ttfamily,breaklines=true,keepspaces=true]
Orchestrator Planning Prompt
-----------------------------------------------------------------------
You are an Orchestrator Agent coordinating information gathering to improve a Fault Analysis Report. You receive a judge evaluation and decide which helper agents to call and what to ask them.
TASK: Create a focused plan to gather only the information that will directly address metrics scored <=3.
DECISION FRAMEWORK:
  Root Cause <=3: Need diagnostic specificity.
    -> LogAgent for exact component/error pattern in logs.
    -> DomainKnowledgeAgent for component behaviour specs.
    -> TicketSimilarityAgent for historical RCA confirmation.
  Resolution <=3: Need concrete fix details.
    -> DomainKnowledgeAgent for exact parameter names, commands, or configuration values.
    -> TicketSimilarityAgent for specific steps from a proven resolution.
  Coverage <=3: Identify missing dimension (symptom/root cause/fix/preventive).
    -> Query the most relevant agent for that dimension.
  Focus <=3: FA distracted by off-topic evidence.
    -> Re-query with questions specific to the fault's component and symptom.
CONSTRAINTS:
  - 2-4 queries total; more queries add noise without proportional benefit.
  - Avoid asking multiple agents for the same information.
  - Do not re-submit a question already asked in a prior round.
Return JSON:
{  "revision_needed":  true | false,
  "reasoning":        "brief explanation of the query plan",
  "agent_queries": [
    { "agent_name": "...", "query": "...", "purpose": "which metric this addresses" }],
  "expected_outcome": "what these queries should add to the report"}
\end{lstlisting}

\section{Evaluation}
\label{sec:eval}

\begin{figure}[!b]
\centering

\begin{minipage}[t]{0.50\textwidth}
\vspace{0pt}
\captionsetup{type=table}
\caption{Evaluation results across 987 fault analysis reports (final refinement round, 0--5 scale). \emph{Overall} is the 5-metric average.}
\label{tab:overall}
\centering
\small
\setlength{\tabcolsep}{3pt}
\renewcommand{\arraystretch}{1.08}

\begin{tabular*}{\linewidth}{@{\extracolsep{\fill}}lccc@{}}
\toprule
Metric & Mean$\pm$Std & Med. & $\geq$3(\%) \\
\midrule
Accuracy        & 2.63$\pm$0.90 & 3   & 59.9 \\
Completeness    & 4.37$\pm$0.70 & 4   & 98.4 \\
Relevance       & 4.88$\pm$0.59 & 5   & 97.9 \\
Clarity         & 4.94$\pm$0.24 & 5   & 100.0 \\
Tech. Depth     & 3.83$\pm$0.85 & 4   & 92.0 \\
\midrule
\textbf{Overall} & \textbf{4.13$\pm$0.43} & 4.2 & 98.4 \\
\bottomrule
\end{tabular*}
\end{minipage}
\hfill
\begin{minipage}[t]{0.46\textwidth}
\vspace{0pt}
\centering
\begin{tikzpicture}
\begin{axis}[
    ybar,
    width=\linewidth,
    height=4.2cm,
    bar width=13pt,
    ylabel={Count},
    xlabel={Accuracy score},
    xtick={0,1,2,3,4,5},
    ymin=0, ymax=520,
    enlarge x limits=0.04,
    nodes near coords,
    nodes near coords align={vertical},
    every node near coord/.append style={font=\scriptsize},
    ymajorgrids=true,
    grid style={dashed,gray!40},
]
\addplot[fill=blue!45!white, draw=blue!70!black] coordinates {
    (0,33) (1,36) (2,327) (3,476) (4,96) (5,19)
};
\end{axis}
\end{tikzpicture}

\captionsetup{type=figure}
\caption{Distribution of Accuracy scores ($n$=987, final refinement round). The distribution is unimodal and left-skewed (skewness=$-$0.44), with the mode at~3.}
\label{fig:acc_dist}
\end{minipage}

\end{figure}
\subsection{Dataset and Setup}

We evaluate ASTRA on 987 real-world telecom fault tickets drawn from a large internal corpus that spans seven product lines covering radio access, baseband, and integrated cell-site equipment. Each ticket carries structured metadata (product, feature, software build), an engineer-written description, and references to system-log files. Ground-truth fault analyses (identification + resolution) are written by domain engineers following a standardized FA template.

 We additionally have access to a \emph{Technical Analysis} (TA) investigation chain: multi-round debugging notes written by domain engineers during live diagnosis, containing raw log snippets, component names, parameter values, and final conclusions. The TA data is used only by the evaluation judge---ASTRA never sees it at inference time---providing a richer reference against which to measure diagnostic depth.

\textbf{Evaluation protocol.}
Each ASTRA-generated report is scored by an LLM judge (GPT-OSS 120B at temperature~0) on five dimensions using a 0--5 Likert scale: \emph{Accuracy}, \emph{Completeness}, \emph{Relevance}, \emph{Clarity}, and \emph{TechnicalDepth} (full evaluation prompt in Appendix~\ref{prompt:eval}). The first four are assessed against the ground-truth FA alone; TechnicalDepth is additionally scored using the TA investigation chain as reference, capturing whether the system provides grounded technical details consistent with the engineers' debugging record. The five scores are averaged into a composite \emph{Overall} score. Scores $\geq 3$ indicate a report that is \emph{useful in practice}: the fault area is correctly identified even if the exact mechanism differs from the ground-truth. The Accuracy rubric uses explicit per-score anchors to reduce scoring ambiguity: score~5 requires the same specific root cause mechanism as the ground-truth; score~3 requires the correct fault area and symptom but allows a different specific mechanism; score~1 requires a wrong fault domain entirely. This hierarchical design follows calibration recommendations for LLM-as-a-Judge systems~\cite{gu_survey_2026,zheng_judging_2023}.

\subsection{Overall Performance}

Table~\ref{tab:overall} reveals a clear two-tier structure. \emph{Presentation quality} is consistently high: Relevance (4.88), Clarity (4.94), and Completeness (4.37) show ceiling effects with 97--100\% of reports exceeding the ``useful'' threshold. TechnicalDepth (3.83) is strong but more variable (92.0\% $\geq$3): as the refinement loop broadens contextual coverage, it sometimes replaces low-level specific details---error codes, parameter values, log identifiers---with more general explanatory content (see Section~\ref{sec:refinement}). This confirms that the claim--evidence architecture reliably produces well-structured, coherent reports. \emph{Accuracy} (2.63) remains the primary bottleneck, indicating that pinpointing the exact causal mechanism---rather than the general fault direction---is the primary challenge. The accuracy distribution (Figure~\ref{fig:acc_dist}) concentrates at score~3 (48.2\% of reports), meaning the system typically identifies the correct symptom and fault area but proposes a different specific mechanism than the ground-truth. Scores~4--5 (exact or near-exact root cause) account for 11.7\%, while catastrophic failures (scores~0--1) remain rare at 7.0\%.

\subsection{Stratification by Fault Type}

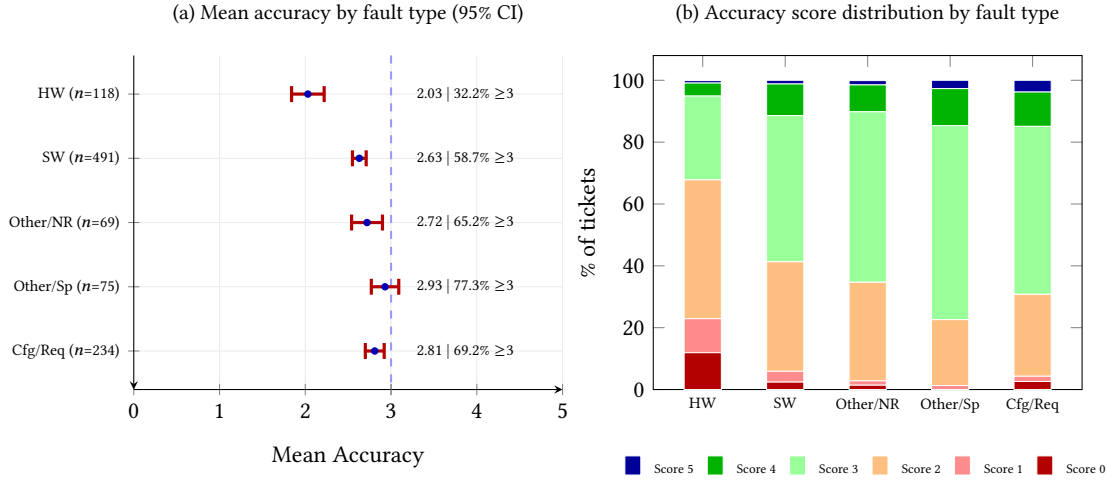
\begin{figure*}[t]
\centering
\begin{tikzpicture}
\begin{axis}[
    name=dotplot,
    width=0.48\textwidth,
    height=6cm,
    xlabel={Mean Accuracy},
    xmin=0, xmax=5,
    xtick={0,1,2,3,4,5},
    ytick={0,1,2,3,4},
    yticklabels={HW ($n$=118), SW ($n$=491), Other/NR ($n$=69), Other/Sp ($n$=75), Cfg/Req ($n$=234)},
    y tick label style={font=\scriptsize},
    y dir=reverse,
    ymin=-0.6, ymax=4.6,
    enlarge y limits=0.12,
    grid=both,
    grid style={gray!15},
    axis lines=left,
    clip=false,
    title={\small (a) Mean accuracy by fault type (95\% CI)},
    title style={at={(0.5,1.02)}},
    extra x ticks={3},
    extra x tick labels={},
    extra x tick style={grid=major, grid style={blue!40, dashed, thick}},
]
\addplot[
    only marks,
    mark=*,
    mark size=1pt,
    blue!70!black,
    thick,
    error bars/.cd,
        x dir=both,
        x explicit,
        error bar style={very thick, red!70!black},
        error mark options={rotate=90, mark size=3pt, very thick},
] coordinates {
    (2.03,0) +- (0.19,0)
    (2.63,1) +- (0.08,0)
    (2.72,2) +- (0.18,0)
    (2.93,3) +- (0.16,0)
    (2.81,4) +- (0.11,0)
};
\node[anchor=west, font=\scriptsize] at (axis cs:3.2,0) {2.03~~|~~32.2\% $\geq$3};
\node[anchor=west, font=\scriptsize] at (axis cs:3.2,1) {2.63~~|~~58.7\% $\geq$3};
\node[anchor=west, font=\scriptsize] at (axis cs:3.2,2) {2.72~~|~~65.2\% $\geq$3};
\node[anchor=west, font=\scriptsize] at (axis cs:3.2,3) {2.93~~|~~77.3\% $\geq$3};
\node[anchor=west, font=\scriptsize] at (axis cs:3.2,4) {2.81~~|~~69.2\% $\geq$3};
\end{axis}

\begin{axis}[
    at={(dotplot.east)},
    anchor=west,
    xshift=1.2cm,
    width=0.48\textwidth,
    height=6cm,
    ybar stacked,
    bar width=14pt,
    ylabel={\% of tickets},
    symbolic x coords={HW, SW, OtherNR, OtherSp, CfgReq},
    xtick=data,
    xticklabels={HW, SW, Other/NR, Other/Sp, Cfg/Req},
    x tick label style={font=\scriptsize},
    ymin=0, ymax=108,
    ytick={0,20,40,60,80,100},
    enlarge x limits=0.15,
    legend style={at={(0.5,-0.18)}, anchor=north, font=\tiny, draw=none, legend columns=6, column sep=3pt},
    title={\small (b) Accuracy score distribution by fault type},
    title style={at={(0.5,1.02)}},
    reverse legend,
]
\addplot[fill=red!70!black, draw=white, line width=0.3pt,
  point meta=explicit symbolic,
  nodes near coords=\pgfplotspointmeta,
  nodes near coords style={font=\tiny, text=white, anchor=center}
] coordinates {
    (HW, 11.9) [] (SW, 2.4) [] (OtherNR, 1.4) [{}] (OtherSp, 0.0) [{}] (CfgReq, 2.6) []
};
\addplot[fill=red!45!white, draw=white, line width=0.3pt,
  point meta=explicit symbolic,
  nodes near coords=\pgfplotspointmeta,
  nodes near coords style={font=\tiny, anchor=center}
] coordinates {
    (HW, 11.0) [] (SW, 3.5) [] (OtherNR, 1.4) [{}] (OtherSp, 1.3) [{}] (CfgReq, 1.7) [{}]
};
\addplot[fill=orange!50!white, draw=white, line width=0.3pt,
  point meta=explicit symbolic,
  nodes near coords=\pgfplotspointmeta,
  nodes near coords style={font=\tiny, anchor=center}
] coordinates {
    (HW, 44.9) [] (SW, 35.4) [] (OtherNR, 31.9) [] (OtherSp, 21.3) [] (CfgReq, 26.5) []
};
\addplot[fill=green!40!white, draw=white, line width=0.3pt,
  point meta=explicit symbolic,
  nodes near coords=\pgfplotspointmeta,
  nodes near coords style={font=\tiny, anchor=center}
] coordinates {
    (HW, 27.1) [] (SW, 47.3) [] (OtherNR, 55.1) [] (OtherSp, 62.7) [] (CfgReq, 54.3) []
};
\addplot[fill=green!70!black, draw=white, line width=0.3pt,
  point meta=explicit symbolic,
  nodes near coords=\pgfplotspointmeta,
  nodes near coords style={font=\tiny, text=white, anchor=center}
] coordinates {
    (HW, 4.2) [{}] (SW, 10.2) [] (OtherNR, 8.7) [] (OtherSp, 12.0) [] (CfgReq, 11.1) []
};
\addplot[fill=blue!60!black, draw=white, line width=0.3pt,
  point meta=explicit symbolic,
  nodes near coords=\pgfplotspointmeta,
  nodes near coords style={font=\tiny, text=white, anchor=center}
] coordinates {
    (HW, 0.8) [{}] (SW, 1.2) [] (OtherNR, 1.4) [{}] (OtherSp, 2.7) [{}] (CfgReq, 3.8) []
};
\legend{Score 0, Score 1, Score 2, Score 3, Score 4, Score 5}
\end{axis}
\end{tikzpicture}
\caption{Accuracy stratified by fault type. (a)~Horizontal dot plot with 95\% confidence intervals  ; the dashed line marks the ``useful'' threshold (score~3). Hardware faults have the lowest mean (2.03) and the widest CI, with only 32.2\% scoring $\geq$3. (b)~Stacked bars show the proportion of tickets in each accuracy bucket. Nearly 68\% of hardware tickets score $\leq$2 (red--orange), compared to $\sim$31\% for Config/Requirement and $\sim$41\% for Software; hardware has almost no score-5 tickets.}
\label{fig:fault_type}
\end{figure*}

We classify tickets into five fault-type categories based on the ground-truth root cause label and examine where ASTRA succeeds and where it struggles. Figure~\ref{fig:fault_type}a reveals that Hardware is the only fault type whose median falls below the ``useful'' threshold: 67.8\% of HW tickets score $\leq$2, compared to 41.3\% for Software and 30.8\% for Config/Requirement. The effect is large (Cohen's $d$=0.80 between HW and Cfg/Req; Kruskal--Wallis $H$=64.2, $p$$<$0.001 across all five fault types). Figure~\ref{fig:fault_type}b makes the distributional difference vivid: the Hardware bar is dominated by red and orange (scores 0--2), while the other categories are dominated by green (score~3). Other/Specialized and Config/Requirement faults achieve the highest proportion of scores~4--5 ($\sim$14--15\%), consistent with these faults being explicitly documented in ticket narratives. The gap is specifically in diagnosing the hardware-level root cause, which requires board-level telemetry and hardware test data that ASTRA cannot access.

\subsection{Root cause category analysis.}
\begin{figure}[t]
\centering

\makebox[\columnwidth][c]{\hspace*{-0.45cm}%
\begin{tikzpicture}
\begin{axis}[
    xbar,
    width=0.86\columnwidth,
    height=5.7cm, 
    bar width=6pt, 
    xlabel={Mean Accuracy (0--5 scale)},
    xmin=0, xmax=3.85,
    xtick={0,1,2,3},
    ytick={0,1,2,3,4,5,6,7,8,9},
    yticklabels={
        3rd-party HW/SW ($n$=106),
        Config.\ error ($n$=45),
        Impl./Coding ($n$=298),
        Design error ($n$=89),
        HW failure ($n$=24),
        Other/Undef.\ ($n$=41),
        Knowledge gap ($n$=28),
        Integr./Design ($n$=113),
        Other ($n$=86),
        Missing Req./Spec.\ ($n$=157)
    },
    y tick label style={font=\scriptsize, align=right, text width=3.5cm},
    enlarge y limits=0.02, 
    extra x ticks={3},
    extra x tick labels={},
    extra x tick style={grid=major, grid style={blue!40, dashed, thick}},
    ymajorgrids=true,
    grid style={dashed, gray!20},
    clip=false,
    nodes near coords={\pgfmathprintnumber[fixed,precision=2]{\pgfplotspointmeta}},
    nodes near coords style={font=\scriptsize, anchor=west, xshift=4pt},
    point meta=x,
]
\addplot[
    fill=blue!40!white,
    draw=blue!65!black,
    error bars/.cd,
        x dir=both,
        x explicit,
        error bar style={thick, gray!60},
        error mark options={rotate=90, mark size=2.5pt, thick},
] coordinates {
    (2.00,0) +- (0.20,0)
    (2.56,1) +- (0.30,0)
    (2.58,2) +- (0.10,0)
    (2.64,3) +- (0.17,0)
    (2.67,4) +- (0.32,0)
    (2.68,5) +- (0.24,0)
    (2.75,6) +- (0.27,0)
    (2.80,7) +- (0.14,0)
    (2.81,8) +- (0.18,0)
    (2.90,9) +- (0.12,0)
};
\end{axis}
\end{tikzpicture}%
}

\caption{Mean Accuracy by ground-truth root cause category, sorted by mean accuracy (final refinement round, 95\% CI error bars). The dashed line marks the ``useful'' threshold (score~3). Faults attributable to 3rd-party hardware/software are hardest (Cohen's $d$=$-$0.74 vs.\ all other categories, $p$$<$0.001); missing requirement/specification faults are most diagnosable, as their causal evidence appears directly in ticket narratives and documentation. Kruskal--Wallis $H$=65.1, $p$$<$0.001.}
\label{fig:rc_acc}
\end{figure}
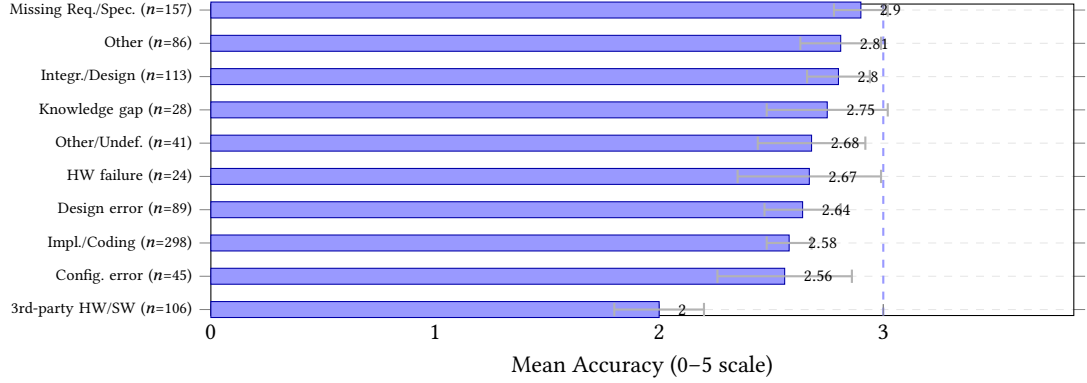

To complement the coarse fault-type view, we further stratify by the ground-truth \emph{root cause category} recorded in each ticket (Figure~\ref{fig:rc_acc}). Ten categories emerge after grouping the 30+ raw labels. Kruskal--Wallis across categories is highly significant ($H$=65.1, $p$$<$0.001), confirming that root cause type is a strong predictor of diagnosability.

Two poles stand out. \emph{3rd-party HW/SW} is by far the hardest category: mean accuracy 2.00, only 33.0\% $\geq$3, Cohen's $d$=$-$0.74 vs.\ all other categories ($p$$<$0.001). These tickets---103 of the 106 are Hardware-typed---involve root causes outside the operator's own codebase, so neither historical ticket precedents, nor log patterns, nor documentation provide reliable diagnostic signals; the system can identify the symptom domain but cannot pinpoint the third-party source. Conversely, \emph{Missing Req./Spec.} tickets are the most diagnosable (Acc=2.90, 72.0\% $\geq$3): these faults stem from explicit gaps in documented specifications, which are directly captured in ticket narratives and documentation---exactly the evidence channels ASTRA is designed to exploit. Integration/design faults (Acc=2.80) follow a similar logic: the fault manifests as a cross-component inconsistency that is often traceable through log sequences and similar historical tickets.
This consistency confirms that the difficulty gradient is structural---driven by information availability in the evidence channels---rather than a scoring artifact.

\subsection{Error Analysis}

To characterize failure modes, we analyze the 396 reports with Accuracy~$\leq$2 (40.1\% of all reports). Categories are derived from accuracy score levels and qualitative inspection are non-exclusive; counts in Table~\ref{tab:errors} reflect the relative frequency of each failure pattern.
The dominant failure mode (Table~\ref{tab:errors}) is a \emph{plausible but incorrect component attribution} (271 tickets, 68.4\%). In these cases the system identifies a related component in the correct product domain---consistent with correct fault-area identification---but not the true responsible unit. The evidence channels surface enough signal to localize the general fault area but not enough to pinpoint the specific component or mechanism, a structurally hard problem given the information asymmetry between ASTRA's evidence (logs, similar tickets, documentation) and the live debugging sessions available to engineers.
\begin{table}[t]
\centering
\small
\caption{Error taxonomy for low-accuracy predictions (Accuracy~$\leq$~2, $n$=396, 40.1\% of reports). Categories are non-exclusive. }
\label{tab:errors}
\begin{tabular}{l r r}
\toprule
Failure Mode & Count & \% \\
\midrule
Plausible but incorrect component & 271 & 68.4 \\
Wrong component or subsystem      & 108 & 27.3 \\
Wrong fault domain (Acc$\leq$1)          & 69  & 17.4 \\
Generic/boilerplate analysis             & 22  &  5.5 \\
Fabricated technical details             & $<$10 & $<$3.0 \\
\bottomrule
\end{tabular}
\end{table}

\emph{Completely wrong component attribution} affects 27.3\% of error cases, and 17.4\% involve the fault being mis-classified at the domain level (Accuracy~$\leq$1). Generic or boilerplate analysis, where the report could apply to any ticket of the same class, accounts for 5.5\%. Critically, \emph{fabricated technical details remain rare} ($<$3\% of low-accuracy cases), suggesting that the claim--evidence architecture effectively constrains confabulation even when the diagnostic conclusion is incorrect.

\subsection{What Drives Accuracy}

We compute the Pearson correlation of each metric with Accuracy ($n$=987). Among the five evaluation metrics, Completeness shows the strongest association ($r$=0.32), followed by Relevance ($r$=0.28), Clarity ($r$=0.22), and TechnicalDepth ($r$=0.15). All four are weak to moderate, confirming that the evaluation dimensions capture largely independent quality aspects rather than redundantly measuring a single construct. This separation also explains why the Overall score remains high even when Accuracy is modest---the presentation metrics, which consistently approach their ceiling, dominate the composite. 
\textbf{Report length is not a proxy for quality.} Longer final reports are \emph{not} more accurate. Dividing reports into quintiles by character count shows a slight negative trend in mean Accuracy (Q1:~2.75, Q5:~2.59), indicating that the system does not compensate for diagnostic uncertainty by producing more text. This is an important sanity check on the refinement loop: the 19\% length growth across rounds reflects the accumulation of new evidence, not stylistic padding, and accuracy gains (where they occur) come from evidence specificity rather than verbosity.

\begin{table}[t]
\centering
\small
\caption{Mean evaluation scores by refinement round (paired analysis, $N$=674 tickets that completed all five rounds).}
\label{tab:rounds}
\begin{tabular}{c r c c c c c c}
\toprule
Round & $N$ & Acc & Comp & Rel & Clar & TD & Overall \\
\midrule
R0 & 674 & 2.72 & 4.26 & 4.94 & 4.91 & 4.36 & 4.24 \\
R1 & 674 & 2.65 & 4.34 & 4.90 & 4.93 & 3.86 & 4.14 \\
R2 & 674 & 2.64 & 4.37 & 4.90 & 4.94 & 3.91 & 4.15 \\
R3 & 674 & 2.62 & 4.36 & 4.86 & 4.94 & 3.90 & 4.14 \\
R4 & 674 & 2.63 & 4.36 & 4.84 & 4.93 & 3.83 & 4.12 \\
\bottomrule
\end{tabular}
\end{table}

\subsection{Refinement Loop Effectiveness}
\label{sec:refinement}

ASTRA's judge--orchestrator loop runs for up to five rounds per ticket. To enable a clean paired comparison, Table~\ref{tab:rounds} reports mean scores across the 674 tickets that completed all five rounds, so the same ticket population is tracked from R0 through R4.
An important interpretive note: the pipeline judge's criteria measure \emph{structural completeness}---whether the report covers all expected dimensions with grounded identifiers---not diagnostic correctness. The loop is therefore best understood as a coverage-refinement mechanism, not an accuracy-improvement mechanism.
The most striking effect of refinement is on TechnicalDepth. The paired R0$\rightarrow$R4 mean delta is $\Delta=-0.53$ ($p$$<$0.001), with 47.2\% of tickets losing technical depth and only 10.5\% gaining. This reflects a systematic side effect: as the judge directs the orchestrator toward broader contextual elaboration to fill coverage gaps, the LLM tends to replace low-level specific details---error codes, parameter values, log-identified identifiers---with more general explanatory content. Completeness modestly improves ($\Delta=+0.10$, 27.4\% improve, 21.5\% degrade, $p$$<$0.001), confirming that the loop achieves its structural objective---broader coverage---but at a cost to technical grounding. Accuracy is largely unchanged ($\Delta=-0.09$, 16.2\% improve, 24.8\% degrade, 59.1\% unchanged, $p$$<$0.001), which is the expected outcome: once the available evidence is exhausted, further iterations can only rephrase and expand; they cannot supply missing diagnostic signal. Relevance ($\Delta=-0.10$) and Clarity ($\Delta=+0.02$) remain near-ceiling throughout, confirming that the refinement loop does not disturb these scores.
Of the 987 tickets, 6.8\% converge at round~0 (the judge endorses the initial draft immediately), while 68.3\% exhaust all rounds. This reveals a natural difficulty gradient: tickets with strong corroborating signals across evidence channels converge quickly; diagnostically ambiguous tickets iterate without necessarily achieving higher accuracy.

\section{Conclusion}

We proposed ASTRA, a multi-agent system for automated fault analysis that coordinates specialized agents over heterogeneous evidence channels---historical tickets, system logs, and technical documentation---through a structured claim--evidence intermediate representation and an iterative judge--orchestrator refinement loop. Evaluated on 987 real-world telecom fault tickets, ASTRA achieves a mean quality score of 4.13/5.0 across five evaluation dimensions, with 59.9\% of reports correctly identifying the fault area. Presentation quality is consistently high (Relevance~4.88, Clarity~4.94), and the claim--evidence architecture limits fabricated technical details to under 3\% of error cases. Stratification by fault type reveals that configuration and software faults are substantially more amenable to automated analysis than hardware faults ($d$=0.80), pointing to a fundamental limitation of text-based evidence channels for hardware fault diagnosis. The dominant failure mode---plausible but incorrect component attribution (68.4\% of Accuracy~$\leq$~2 cases)---highlights an inherent ceiling: resolving root cause ambiguity often requires the kind of live debugging and internal knowledge transfer that ASTRA's evidence channels cannot yet provide.

\section{Limitations}

Several limitations should be noted. First, evaluation relies on a single LLM judge (GPT-OSS 120B); despite explicit per-score anchors and zero temperature, judge scores may not perfectly align with expert human judgment. Validating against engineer ratings is an important next step. Second, the 2,000-line cap on diagnostic log extraction constrains the LogAgent; 84\% of tickets saturate this cap, which may limit the system's ability to surface rare diagnostic signals buried deeper in the logs. Third, ASTRA has no access to hardware-level diagnostic data---board sensor readings, hardware event logs, or physical component test results---which limits effectiveness on hardware faults, the category with the lowest accuracy (2.03). Fourth, the evaluation covers a single organization's ticket corpus; generalization to other operators or domains requires further validation. 

\section{Future Work}

Several directions emerge from this work. \emph{Adaptive log budgets}: dynamically adjusting the log extraction cap based on ticket complexity and initial signal strength could improve accuracy for the 68.3\% of tickets that currently exhaust all refinement rounds. \emph{Refinement-aware training}: the refinement loop currently improves coverage (Completeness $+$0.10) but degrades technical specificity (TechnicalDepth $-$0.54); training the report generator to preserve evidence-grounded details across iterations could recover this trade-off. \emph{Hardware-aware agents}: integrating hardware-level diagnostic data (board sensor readings, hardware event logs, component test results) as an additional evidence channel could address the domain-level misclassifications that affect 17.4\% of low-accuracy cases. \emph{Human-in-the-loop evaluation}: conducting a controlled study with domain engineers to validate LLM-based scoring against expert judgment and measure time savings when engineers use ASTRA reports as starting points. \emph{Retrieval-aware training}: fine-tuning the ticket similarity model to optimize for root cause category overlap, rather than surface-level symptom similarity, could reduce the dominant ``plausible but incorrect component'' failure mode.

\section*{Acknowledgments}
We thank Gabriel G\'orski, Jakub Kozerski, Bartlomiej Ruszaj, and Adrian Dudycz from Nokia Mobile Infrastructure for collecting the telecom-related troubleshooting tickets, fault analyses, technical analyses, and logs. We thank Johann Daigremont from Nokia Bell Labs for creating the MCP server used by the DomainKnowledgeAgent. We thank Ahmet Akyamac from Nokia Bell Labs for collecting the internal acronyms used to augment tickets during the offline preprocessing step before executing the agentic workflow. We thank our past summer intern Yuting Hu for implementing the first version of the agentic workflow using Nokia's ticket troubleshooting data.



\newpage
\bibliographystyle{ACM-Reference-Format}
\bibliography{main}

\newpage
\appendix

\section{ASTRA Prompts}
\label{appendix:prompts}

This appendix reproduces the system-level instruction prompts from ASTRA's pipeline. The claim extraction, evidence mapping, and orchestrator planning prompts also appear inline in Section~4 alongside the corresponding pipeline descriptions. All LLM calls use \texttt{gpt-oss:120b} at temperature~0 unless otherwise noted. Prompts marked with variable sections (e.g., ticket context, log content, agent outputs) show the static instruction portions; dynamic content is assembled at runtime by the orchestration code.

\subsection{TicketSimilarityAgent Analysis Prompts}
\label{prompt:ticketsimilarity}

The TicketSimilarityAgent uses a two-phase LLM reranker to select the most diagnostically useful historical tickets from candidates retrieved by dense vector search. Phase~1 processes batches of 20 candidates each; Phase~2 makes the final selection from the Phase~1 survivors.

\textbf{Phase~1 Batch Reranker System Message:}

\begin{lstlisting}[basicstyle=\footnotesize\ttfamily,breaklines=true,keepspaces=true]
You are a Nokia 5G/LTE RAN fault analyst reviewing historical tickets.

A ticket is USEFUL for the target if reading its Identification and Resolution would meaningfully help an engineer diagnose or fix the target fault. Concretely:
  - The root cause operates in the same SW component or subsystem
  - The failure mechanism (what broke and why) is directly comparable
  - The resolution gives actionable information: a specific parameter, flag, function, or configuration change relevant to the target

A ticket is NOT useful if:
  - Its root cause is in a different subsystem (different SW layer, different feature)
  - The root cause is blank, "not defined", "unknown", or too vague
  - The only connection is a shared surface keyword with nothing transferable about how or why it failed

Note: embedding similarity is shown for context only -- it does not indicate usefulness.

Respond ONLY with a valid JSON object:
{"selected": [{"pr_id": "<id>", "usefulness": "HIGH|MEDIUM|LOW",
  "reason": "<one sentence>"}],
 "excluded_reason": "<one sentence>"}
Select at most 5. Fewer or empty list is valid.
\end{lstlisting}

\textbf{Phase~2 Final Selection System Message:}

\begin{lstlisting}[basicstyle=\footnotesize\ttfamily,breaklines=true,keepspaces=true]
You are a senior Nokia 5G/LTE RAN fault analyst making the final selection of historical tickets to anchor a root cause analysis.

Apply the same usefulness test as Phase 1. Additional guidance:
  - Prefer tickets with specific, confirmed root causes over vague ones
  - If several tickets describe the exact same fault mechanism, pick the one with the most complete and actionable resolution; exclude the rest as duplicates
  - Diverse coverage is better than five tickets all pointing to the same hypothesis
  - A wrong-domain ticket in the final 5 actively harms downstream analysis -- when in doubt, exclude it

You may agree or override Phase 1 usefulness ratings.

Respond ONLY with a valid JSON object:
{"final_selection": [{"pr_id": "<id>", "rank": <1-5>,
  "usefulness": "HIGH|MEDIUM|LOW", "reason": "<1-2 sentences>"}],
 "excluded": {"<pr_id>": "<one-sentence reason>"},
 "summary": "<2 sentences: what fault domain do the selected tickets collectively point to>"}
Rank 1 = most useful. At most 5.
\end{lstlisting}

Both phases receive a user message containing the target ticket metadata (product, feature, build, title, description) and the candidate block with ticket details and similarity scores.

\subsection{LogAgent Analysis Prompts}
\label{prompt:logdigest}

The LogAgent issues a single constrained LLM call. The prompt template assembles: ticket context, enhanced description, acronym expansions, component context, an optional similar-ticket reference block (shown below), and the filtered log content. The system message and appended analysis task are:

\begin{lstlisting}[basicstyle=\footnotesize\ttfamily,breaklines=true,keepspaces=true]
LogAgent Prompt
-----------------------------------------------------------------------
System message:
  You are a precise log evidence extractor for wireless network equipment.
  You ONLY report what you can directly see in the provided log lines.
  You NEVER invent details not present in the logs.
  When in doubt, say 'not determinable from logs'.

Analysis task (appended after log lines):
  Analyze ONLY the log lines provided above. Extract factual observations.

  STRICT RULES:
  1. EVERY finding must directly quote at least one log line from above.
  2. DO NOT invent file paths, function names, or error codes not in the logs.
  3. DO NOT speculate about code-level root causes unless the log says so.
  4. If root cause is unclear, say "Root cause not determinable from logs alone".
  5. Focus on WHAT happened (observable facts) not WHY (speculation).

  Return JSON:
  {
    "findings": [ { "type", "component", "log_quote", "timestamp",
                    "frequency", "severity", "description",
                    "fault_relevance", "relevance_reasoning" } ],
    "temporal_sequence": "...",
    "root_cause_assessment": {
      "from_logs": "...", "confidence": "high|medium|low|none",
      "needs_further_investigation": "..."
    },
    "summary": "2-3 sentence summary based ONLY on log evidence"
  }
\end{lstlisting}

When similar tickets are available from the pre-fetch, the following reference block is prepended to the log lines section:

\begin{lstlisting}[basicstyle=\footnotesize\ttfamily,breaklines=true,keepspaces=true]
Similar Tickets Reference Block
-----------------------------------------------------------------------
SIMILAR RESOLVED HISTORICAL TICKETS [LOG INTERPRETATION GUIDE]
Use these confirmed resolutions to:
  1. Classify log findings: same component/error pattern = likely CAUSAL
  2. Identify ambiguous log token owners
  3. Confirm root cause plausibility -- if logs match the reference pattern, say so
Do NOT copy the resolution verbatim. Only include claims corroborated by log lines above.

  [HIGH/GOOD/POSSIBLE MATCH] Reference N: <ticket title>
  Root cause: <root cause text up to 200 chars>
  Resolution: <resolution summary up to 300 chars>
  Investigation trail: <TA trail for top-2 ranked tickets, up to 400 chars>
\end{lstlisting}

\subsection{DomainKnowledgeAgent Analysis Prompts}
\label{prompt:domainknowledge}

The DomainKnowledgeAgent queries technical documentation via MCP (Model Context Protocol). It uses a system message that constrains the response to fault-relevant content, and an enriched task query constructed from the ticket context and other agents' outputs.

\textbf{System Message:}

\begin{lstlisting}[basicstyle=\footnotesize\ttfamily,breaklines=true,keepspaces=true]
You are a technical documentation assistant for telecom network equipment.
Focus your answers on fault-relevant content: root cause explanations, error code meanings, configuration parameters, and resolution procedures.
Cite the specific document or section when available.
If no retrieved document directly addresses the query, state that clearly rather than generalizing.
Structure your response to distinguish between identification-relevant findings (symptoms, error meanings, diagnostic procedures) and resolution-relevant findings (fixes, configuration changes, workarounds).
\end{lstlisting}

\textbf{Enriched Task Query (constructed at runtime):}

\begin{lstlisting}[basicstyle=\footnotesize\ttfamily,breaklines=true,keepspaces=true]
Find documentation for this Nokia fault -- root cause, known issues, and resolution: {ticket_title} [Alarm IDs: {extracted_FIDs}].
Focus on: (1) what triggers this alarm, (2) configuration parameters and correct values, (3) step-by-step resolution procedure.

CONTEXT FROM OTHER AGENTS (use to narrow your documentation search):
Log analysis findings:
{logagent_output[:2000]}

Similar historical tickets (likely fault domain):
{tsa_output[:1500]}
\end{lstlisting}

The agent output context is truncated (2,000 characters from LogAgent, 1,500 from TicketSimilarityAgent) to stay within the documentation retrieval system's query limits. 

\subsection{Report Generation Prompt}
\label{prompt:report}

This prompt is used for both the initial report and each revision pass. During revision, the judge's feedback and new agent findings from the follow-up round are appended before the claim--evidence block.

\begin{lstlisting}[basicstyle=\footnotesize\ttfamily,breaklines=true,keepspaces=true]
Report Generation Prompt
-----------------------------------------------------------------------
You are a telecom fault analysis expert. Generate a fault analysis report using the provided claim-evidence mapping.

The report has exactly two sections:

identification -- How was the fault detected and diagnosed?
  Required: initial alarm/symptom -> diagnostic steps -> causal chain -> root cause.
  Include: FID/alarm IDs, exact error messages, configuration parameters, log citations, and the name of the affected module or protocol layer.

resolution -- How was the fault resolved?
  Required: exact fix applied -> specific parameter values changed -> verification.
  If a workaround was applied instead of a permanent fix, state that explicitly.

Source attribution:
  For every technical claim, state which agent provided it.
  [VERIFIED] claims are directly grounded -- cite as facts.
  [UNVERIFIED] claims: prefix with "Analysis suggests (unverified)..." -- do NOT state specific IDs or values from unverified claims as confirmed facts.

Quality constraints:
  - 300-600 words per section; completeness takes priority over length targets.
  - Do NOT repeat information -- if LogAgent and TSA both describe the same root cause, merge them, citing both.
  - NEVER invent specific values (IDs, file paths, parameters, timestamps) absent from the evidence.

TicketSimilarityAgent conflict rule:
  If TSA matches describe a different fault domain than LogAgent findings, do NOT follow the TSA match. State the conflict and give log evidence priority.

Return JSON: { "identification": "...", "resolution": "..." }
\end{lstlisting}

\subsection{Judge Evaluation Prompt}
\label{prompt:judge}

The judge evaluates each generated report on a 1--5 scale across five criteria. A score $\leq$3 on any criterion sets \texttt{needs\_revision: true} and triggers a follow-up round.

\begin{lstlisting}[basicstyle=\footnotesize\ttfamily,breaklines=true,keepspaces=true]
Judge Evaluation Prompt
-----------------------------------------------------------------------
You are a Quality Assurance Judge evaluating a Fault Analysis Report (FA) for technical accuracy and factual grounding. You are given: (1) agent outputs, (2) the claim-evidence mapping, (3) the FA report, (4) the original ticket context.

Goal: verify the FA is grounded in the evidence and identify improvements.
You do NOT have access to ground truth -- evaluate only against what agents found.

METRIC 1: Root Cause Identification
  5 -- Precise root cause (specific component + failure mode), corroborated by >=2 sources.
  4 -- Specific root cause, supported by one strong agent source.
  3 -- General fault area described; exact mechanism not pinpointed from evidence.
  2 -- Vague throughout, or names component absent from all agent outputs.
  1 -- No root cause, or directly contradicts agent reports.
  Red flag (score <=2): FA asserts root cause mechanism untraceable to any agent output.
METRIC 2: Resolution Specificity
  5 -- Every step cites an agent source with exact identifiers.
  4 -- Most steps specific; at most one generic step.
  3 -- Correct fix type described but vaguely -- missing specific parameters/values.
  2 -- Generic advice with no ticket-specific details.
  1 -- Recommends actions absent from or contradicted by all agent outputs.
METRIC 3: Evidence Grounding
  5 -- >=90% of specific identifiers appear verbatim in agent outputs.
  4 -- 70-90% grounded; minor unverified claims clearly marked.
  3 -- 50-70% grounded -- several non-trivial claims lack direct evidence.
  2 -- <50% grounded.
  1 -- FA invents specific identifiers not present in any agent output.
METRIC 4: Coverage & Completeness
  5 -- All four dimensions (symptom, root cause, fix, verification) with evidence.
  4 -- All four present; one has limited depth.
  3 -- Three dimensions present; one missing or placeholder.
  2 -- Only two dimensions addressed.
  1 -- Only symptom description; root cause and resolution absent.
METRIC 5: Focus & Relevance
  5 -- Every clause directly addresses the ticket fault.
  4 -- Mostly on-topic; one or two minor tangents.
  3 -- Partially distracted by off-topic evidence.
  2 -- Significant portion (>=1 paragraph) built around unrelated evidence.
  1 -- Narrative anchored to evidence unrelated to reported symptoms.

For any metric scoring <=3, set improvement_needed: true and specify: suggested_focus: what is missing or could be made more specific recommended_agents: which agent is most likely to supply the missing evidence
Set needs_revision: true if ANY metric scores <=3.
Set revision_priority to the single most impactful metric to strengthen first.

Return a single valid JSON object with all scores and improvement recommendations.
\end{lstlisting}

\subsection{TA-Aware Evaluation Prompt}
\label{prompt:eval}

This prompt is used by the external evaluation judge (Section~\ref{sec:eval}) to score each final ASTRA-generated report against both the ground-truth FA and the engineer TA investigation chain. It produces the five reported metric scores in Table~\ref{tab:overall} which are Accuracy, Completeness, Relevance, Clarity, and TechnicalDepth. The evaluation judge is distinct from the pipeline JudgeAgent (Appendix~\ref{prompt:judge}): the pipeline judge scores against \emph{agent outputs} during generation to trigger refinement, whereas this evaluation judge scores against \emph{engineer references} to assess final quality.

\begin{lstlisting}[basicstyle=\footnotesize\ttfamily,breaklines=true,keepspaces=true]
TA-Aware Evaluation Judge Prompt
-----------------------------------------------------------------------
You are evaluating a predicted fault analysis (FA) report for a telecom software defect ticket. You have TWO reference sources:
  1. The official Ground Truth FA (concise engineer summary)
  2. The Technical Analysis (TA) investigation chain (multi-round engineer debugging notes)

IMPORTANT CONTEXT: The predicted FA was generated by an automated AI system that works from log files, similar historical tickets, and technical documentation. It did NOT have access to the TA investigation chain or the live debugging sessions that engineers performed. Evaluate what the system achieves from its available evidence sources -- do not penalize it for lacking information that is only obtainable through live debugging or internal team hand-offs.

--- TICKET INFORMATION ---
<ticket metadata and description>

--- GROUND TRUTH FAULT ANALYSIS (official FA template) ---
Concise engineer-written summary. Often sparse but captures the final root cause category and fix direction.
<ground truth text>

--- TECHNICAL ANALYSIS CHAIN (engineer investigation notes) ---
Multiple analysis rounds from engineers during live debugging. Contains raw log snippets, specific parameter names, component names, and final conclusions. Use this as supplementary context to better understand the fault -- but remember the AI system never saw this data.
<TA chain text>

--- PREDICTED FAULT ANALYSIS ---
Produced by a multi-agent AI system using log analysis, similar ticket retrieval, and documentation queries. Has two sections: IDENTIFICATION and RESOLUTION.
<generated FA report>

--- EVALUATION ---
Evaluate the predicted FA on these 5 metrics. Score each from 0-5.

1. **Accuracy** -- Does the predicted FA identify the same fault domain as the GT? Evaluate in two layers: (a) symptom/fault-area alignment, then (b) root cause mechanism alignment. Use the TA to understand the full picture, but evaluate primarily based on whether the prediction is directionally consistent.
   - Score 5: Same symptom AND same specific root cause mechanism as GT/TA
   - Score 4: Same symptom AND same root cause direction, minor differences
   - Score 3: Same symptom AND same fault area/component family, but different specific root cause mechanism or code path
   - Score 2: Related product domain but different fault area, OR correct symptom but entirely fabricated cause
   - Score 1: Different fault domain entirely (wrong subsystem, wrong symptom category)
   - Score 0: Completely unrelated fault
2. **Completeness** -- Does the prediction address all four expected fault analysis dimensions: (a) symptom/alarm detection, (b) root cause analysis, (c) resolution/fix steps, (d) verification or preventive measures?
   Do NOT penalize for missing TA-specific details the AI system could not access; evaluate whether the prediction covers the essential diagnostic dimensions from its own evidence.
3. **Relevance** -- Does the prediction stay focused on this specific ticket? No off-topic information.
4. **Clarity** -- Is the predicted FA well-structured and easy to follow for a technical engineer?
5. **TechnicalDepth** -- Does the FA include specific technical details grounded in evidence (parameter names, error codes, log excerpts, alarm IDs, component identifiers, timestamps)? Details that also appear in the TA score highest, but specific details derived from log analysis or similar tickets are also valuable.
   - Score 5: Cites specific identifiers and multiple also appear in TA
   - Score 4: Includes concrete technical details grounded in log/ticket evidence; some align with TA findings
   - Score 3: Mentions the right subsystem and symptom with some specific details from its sources
   - Score 2: Mostly generic but includes one or two specific details
   - Score 1: Generic description only -- could apply to any similar ticket
   - Score 0: Completely generic or fabricated technical details
ACCURACY DECISION RULE: If the prediction correctly identifies the symptom AND the affected subsystem/component family, score Accuracy >= 3 regardless of whether the specific root cause mechanism matches. Score 2 requires the fault AREA (not just the mechanism) to be wrong. Score 1 requires the fault DOMAIN to be wrong.

Key principles:
- The prediction was generated WITHOUT access to the TA chain. Do not require TA-level specificity for scores 3-4.
- A detailed, evidence-grounded prediction that is directionally consistent with GT should score 3-4 on Accuracy even if it reaches a slightly different specific mechanism than TA.
- Reserve scores 0-2 for predictions where the fault domain or component is fundamentally wrong.
- Generic boilerplate answers score low on TechnicalDepth even if directionally correct.

Return a JSON object:
{
  "Accuracy": {"analysis_and_justification": "<explanation>", "score": <0-5>},
  "Completeness": {"analysis_and_justification": "<explanation>", "score": <0-5>},
  "Relevance": {"analysis_and_justification": "<explanation>", "score": <0-5>},
  "Clarity": {"analysis_and_justification": "<explanation>", "score": <0-5>},
  "TechnicalDepth": {"analysis_and_justification": "<explanation>", "score": <0-5>}
}
\end{lstlisting}

\end{document}